\documentclass[reprint,aps,prx,showpacs,amsmath,amssymb,floatfix,groupedaddress,longbibliography]{revtex4-2}
\pdfoutput=1
\usepackage[export]{adjustbox}
\usepackage{graphicx}
\usepackage{amsthm}
\theoremstyle{definition}
\usepackage{dcolumn}
\usepackage{bm}
\usepackage{epstopdf}
\usepackage{braket}
\usepackage{mathtools}
\usepackage[svgnames]{xcolor}
\PassOptionsToPackage{hyphens}{url}
\usepackage{hyperref}
\hypersetup{colorlinks=true, citecolor=NavyBlue, urlcolor=NavyBlue, linkcolor=NavyBlue}
\usepackage[caption=false]{subfig}

\newcommand{\figref}[1]{Fig.~\ref{#1}}

\usepackage{etoolbox}
\apptocmd{\sloppy}{\hbadness 10000\relax}{}{}

\DeclareMathOperator{\Tr}{Tr}

\newtheorem*{defn}{Definition}

\newtheorem*{example}{Example}

\begin{document}

\title{Nonlocal Variable-Strength Measurements of N Qubits\texorpdfstring{\\Using GHZ-like Entanglement}{}}

\author{Pierre Vidil}
\email{pierre@quantum.riec.tohoku.ac.jp}
\author{Keiichi Edamatsu}
\affiliation{Research Institute of Electrical Communication, Tohoku University, Sendai 980-8577, Japan}

\date{\today}

\begin{abstract}
The direct measurement of nonlocal properties of entangled quantum systems has been the subject of several recent experimental investigations.
Of particular interest is the implementation of nonlocal measurements via indirect measurement schemes, which allow for greater flexibility in the control of the measurement strength.
Building on previous results established in the bipartite case, we present a scheme to implement genuine nonlocal measurements of N-qubit systems with variable strength, using GHZ-like entangled qubit meters.
This method can be applied to the joint measurement of commuting product observables, enabling us to distinguish between orthogonal nonlocal states, such as Bell states, with minimal disturbance and arbitrary resolution.
An explicit relation between the overall measurement strength and the meter entanglement as quantified by the $n$-tangle is derived, opening the door to a new interpretation of the $n$-tangle as a resource for nonlocal measurements.
\end{abstract}

\maketitle

\section{Introduction}
\label{sec:introduction}

Measuring physical systems and interpreting results is at the basis of any experimental science.
Unlike in classical physics where the study of the measurement process is often relegated to the second plan, the action of measuring and its impact on a quantum system is an integral part of quantum theory \cite{Preskill}.
The so-called \emph{measurement problem} has been at the core of many foundational debates \cite{Laloe2012}; in the meantime, advances in the description of quantum measurements led to a better understanding of the interplay between open systems, entanglement and parameter estimation theory \cite{Davies1976,Kraus1983,Braginsky1992,Helstrom1976}, which in turn generated various novel methods for analysing and controlling the properties of individual quantum systems \cite{Brune1996,Hatridge2013}.

Defining precisely and finding ways to evaluate quantities such as measurement error, disturbance, back-action and measurement strength is important when designing and evaluating the performance of communication protocols which rely on the measurement of quantum systems.
The study of systems under weak measurement interaction also led to the formalism of weak values \cite{Aharonov1988}, which thereafter was applied to quantum metrology \cite{Hosten2008,Dixon2009} and the study of quantum foundational paradoxes \cite{Aharonov2002,Yokota2009,Lundeen2009}, among other things.

On the other hand, the problem of determining which quantities are measurable for local observers \cite{Aharonov1981,Aharonov1986,Popescu1994} and at what cost \cite{Jozsa2003,Clark2010,Groisman2015}, is also of importance when devising practical implementations of quantum communication protocols between distant parties. 
Indeed, quantum theory allows the existence of nonlocal states, which when distributed between several space-like separated observers, exhibit super-classical correlations \cite{Bell1964,Aspect1981,Horodecki2007}.
These constitute a resource useful e.g. in quantum teleportation \cite{Bouwmeester1997} or device-independent certification \cite{Acin2006}.

Schemes to measure the observables associated with such states, both in destructive and non-destructive manners, have been proposed \cite{Vaidman2003,Brodutch2016,Wu2016} and successfully implemented experimentally \cite{Li2019,Pan2019,Xu2019}.
However, these are often sub-optimal in terms of resources, in that they rely on the use of maximally-entangled meters, even for weak or incomplete measurements.

Following previous works concerning bipartite qubit systems \cite{Edamatsu2016a,Vidil2019}, we present here a novel scheme to measure multipartite nonlocal qubit systems with arbitrary measurement strength, and which is resource-efficient in terms of meter entanglement.
This paper is structured as follows: after introducing the concept of variable-strength measurement in Sec. \ref{sec:vsm}, we explicitly construct a measurement scheme for nonlocal product observables, starting with the 2-outcome case in Sec. \ref{sec:nvsm-1}, and then the many-outcome case in Sec. \ref{sec:nvsm-2}.
Finally in Sec. \ref{sec:tangle}, we explore the relation between measurement strength and meter entanglement, using the $n$-tangle as a measure of entanglement.

\section{Notations and tools}
\label{sec:vsm}

\subsection{PVMs and POVMs}
Let $\mathcal{H}$ be a Hilbert space of dimension $n<\infty$.
The standard approach to quantum measurement is via so-called \emph{projective measurements}, for which the effect of getting a result $a_i$ is represented by the projector $P_i$ on the eigenspace associated with the result, $\rho\rightarrow P_i\rho P_i$.
The different possible post-measurement states, and thus the statistics, are then completely determined by the set of projectors $\{P_i\}$, one for each outcome, satisfying the following completeness and orthogonality conditions:
\begin{equation}
    \sum_i P_i=I \qquad P_iP_j=\delta_{ij}P_i \label{eq:PVM_def}
\end{equation}
These conditions guarantee that the quantities formed by the Born rule $p_i=\Tr(\rho P_i)$ constitute a well-defined probability measure on the outcome space.
Such an orthogonal resolution of identity is called a Projection-Valued Measure (PVM) \cite{Paris2012}.

For any observable $A$, i.e. a self-adjoint operator on $\mathcal{H}$, the spectral theorem guarantees the existence of a unique associated PVM $\mathbf{P}^A\equiv\left\{P^A_i\right\}$ such that
\begin{align}
    A=\sum_ia_iP^A_i \label{eq:spectral_theorem}   
\end{align}
where $a_i$ and $P^A_i$ are respectively the eigenvalue and the projector on the eigenspace associated with the outcome $i$ \cite{Hall2013}. 
Note that the observable-PVM correspondance is many-to-one, which is the reason why in the following we describe strong measurements in terms with PVMs rather than observables.

The conditions \eqref{eq:PVM_def} are sufficient but not necessary to define a probability measure for the measurement results.
In particular, the orthogonality condition between different final states can be discarded and still lead to a well-defined interpretation of the measurement process, should one only use a non-orthogonal resolution of identity instead.
Such measurements are called $\emph{generalized measurements}$ and are characterized by a set of positive operators $\{E_i\}$, one per different outcome, satisfying the condition:
\begin{equation}
    \sum_i E_i=I \qquad E_i\geq 0
\end{equation}
These conditions are in turn necessary and sufficient to define probabilities via $p_i=\Tr(\rho E_i)$, and form what is called a \emph{Positive-Operator-Valued Measure} (POVM) \cite{Heinosaari2011}.
A POVM element $E_i$ is called the \emph{effect} associated to the outcome $i$.

Every generalized measurement described by a POVM can be implemented via a projective measurement in a larger Hilbert space, a construction known as a Naimark extension \cite{He2007}.
In particular, Naimark's theorem guarantees that one can always realize POVMs via indirect measurement models, in which an independently prepared additionnal ancilla state, called the \emph{meter}, is made to unitarily interact with the system so that a projective measurement on the meter alone yields the desired statistics \cite{Naimark1940,Peres1990}.
Given a POVM, the problem of finding a corresponding indirect measurement model is in general not trivial and is paramount to any experimental implementation \cite{Sparaciari2013}.

The present paper aims at explicitly constructing such indirect measurement models for a particular class of generalized measurements, which we call nonlocal variable-strength measurements.

\subsection{Variable-strength measurements}

In the following, we restrict ourselves to a $d$-element PVM $\mathbf{P}=\left\{P_i\right\}$ whose elements $P_i$ all have the same rank $r=n/d$.
We then define the notion central to this paper:
\begin{defn}
    A variable-strength measurement (VSM) of a PVM $\mathbf{P}$ is a generalized measurement described by a POVM $\mathbf{E}$ whose effects have the form
    \begin{align*}
        E_i=f(P_i,\mathfrak{s})
    \end{align*}
    with $\mathfrak{s}\in[0,1]$ and $f$ such that for any outcome $i$: 
    \begin{align*}
        f(P_i,\mathfrak{s})&\xrightarrow[\mathfrak{s}\rightarrow0]{} \frac{I}{d} &\text{(no measurement)}\\
        f(P_i,\mathfrak{s})&\xrightarrow[\mathfrak{s}\rightarrow1]{} P_i &\text{(strong measurement)}
    \end{align*}
    where $I$ is the identity operator on $\mathcal{H}$. The parameter $\mathfrak{s}$ is called the \emph{measurement strength}.
\end{defn}
Intuitively, one may think of a VSM as a process that can resolve between different orthogonal subspaces (the images of the $P_i$) more or less accurately, with a resolution depending on the strength $\mathfrak{s}$.
When $\mathfrak{s}=0$, the outcome probability is independent of the system state and is given by a uniform distribution.

Requiring that $f$ be linear in $\mathfrak{s}$ yields the following useful expressions for the VSM effects:
\begin{align}
    E_i&=\frac{1}{d}\left(I+\mathfrak{s}\left(dP_i-I\right)\right)\nonumber\\
    &=\frac{1}{d}\left(\left(1+\mathfrak{s}(d-1)\right)P_i+\left(1-\mathfrak{s}\right)\sum_{j\neq i}P_j\right) \label{eq:effect_decomposition} 
\end{align} 

\begin{figure}[htp]
    \includegraphics[width=0.45\textwidth]{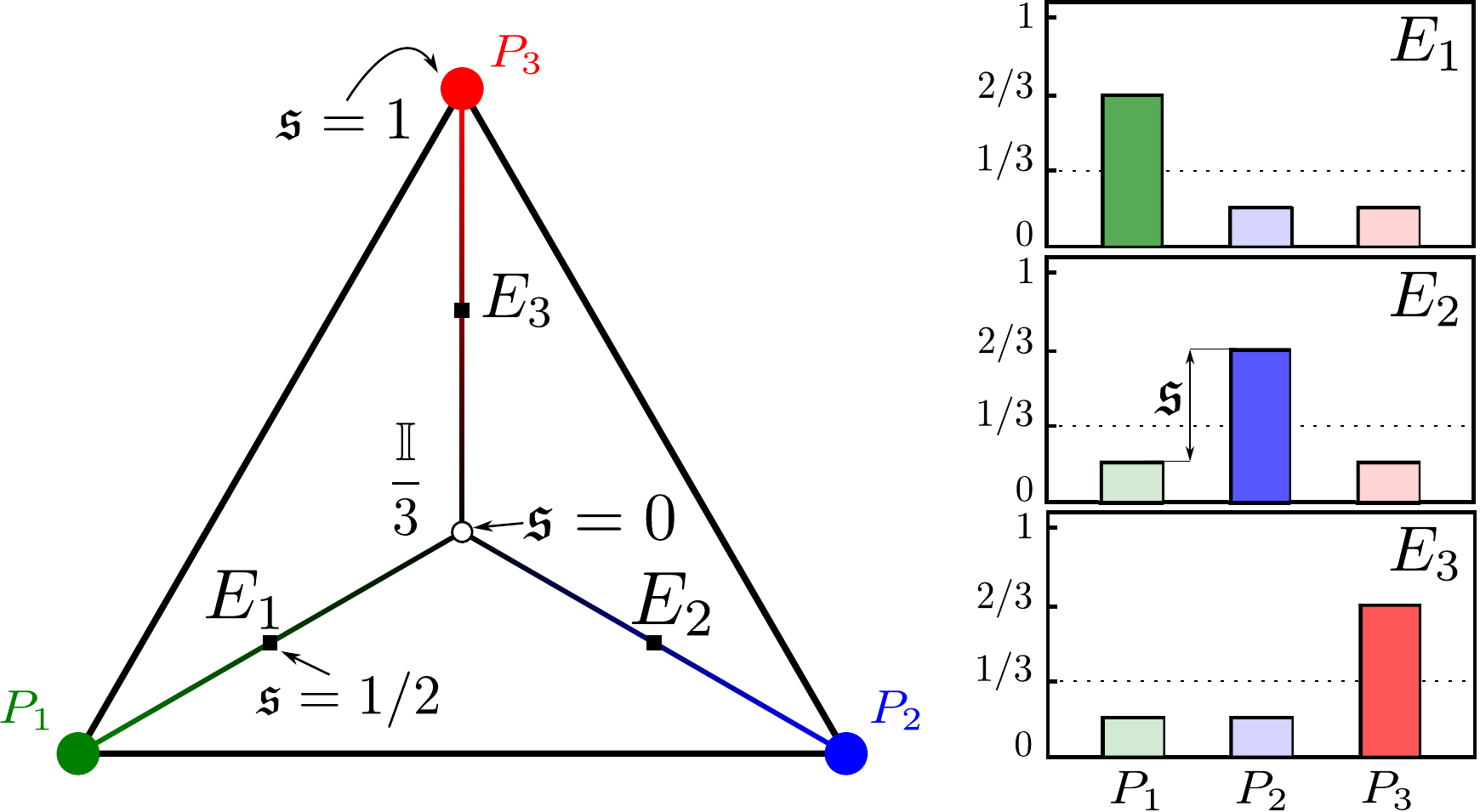}
    \caption{\label{fig:simplex}Graphical representation of a 3-outcome VSM with effects as points in a 2-simplex. The barycenter corresponds to no measurement while the vertices correspond to a strong measurement. 
    The case $\mathfrak{s}=1/2$ is explicited, with the weights for each point depicted on the right.}
\end{figure}

This last decomposition in terms of $d$ coefficients that sum up to one allows us to picture VSM effects as points in a $(d-1)$-simplex, as in \figref{fig:simplex}.
In such a representation, the vertices correspond to the PVM elements while the center corresponds to a uniform result distribution, i.e. no measurement at all.

Equation \eqref{eq:effect_decomposition} also gives a more straightforward interpretation for the measurement strength $\mathfrak{s}$ as the difference between the probability  of getting the correct outcome versus any other arbitrary outcome.
In other words, the measurement strength characterizes the bias of the result probability distribution towards the correct outcome.

\begin{figure}[htp]
    \includegraphics[width=0.40\textwidth]{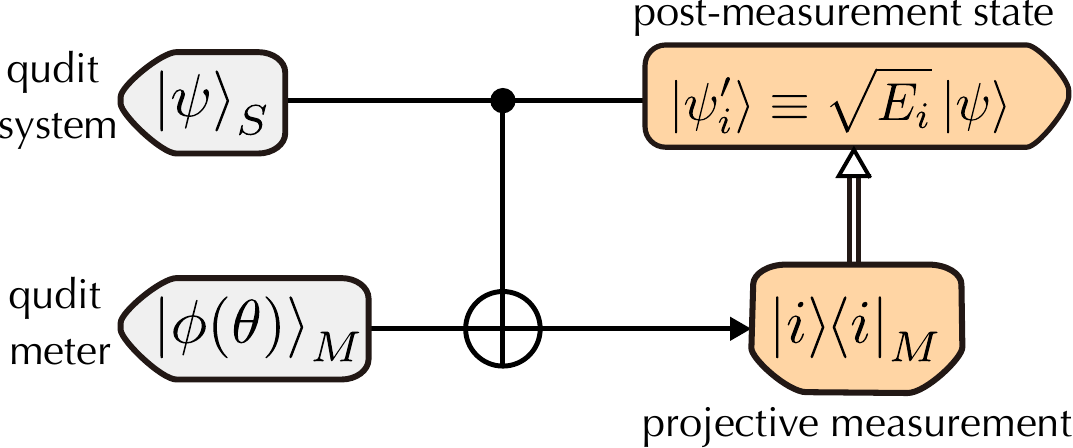}
    \caption{\label{fig:qudit}Schematic representation of an indirect measurement model implementing a VSM on a qudit system using an additionnal qudit meter and a CNOT-like interaction.}
\end{figure}

\begin{example}
    Let $\ket{\psi}_S\in\mathbb{C}^d$ be a qudit system. 
    A VSM of $\left\{\ket{i}\bra{i}_S\right\}_{i=0}^{d-1}$ can be implemented by the indirect measurement model $\left(\mathbb{C}^d,\ket{\phi}_M,U_{CNOT},\left\{\ket{j}\bra{j}_M\right\}_{j=0}^{d-1}\right)$ \cite{Pryde2005, Lund2010, Ho2017} (see \figref{fig:qudit}) with:
    \begin{subequations}
        \begin{align}
            &\ket{\phi(\theta)}_M = \cos\theta \ket{0}_M + \frac{\sin\theta}{\sqrt{d-1}}\sum_{j=1}^{d-1}\ket{j}_M\label{eq:qudit-meter}\\
            &U_{\text{CNOT}}\left(\ket{i}_S\otimes\ket{j}_M\right)= \ket{i}_S\otimes\ket{(i+j) \; \mathrm{mod} \; d}_M \nonumber
        \end{align}
        The POVM effects of the VSM are given by:
        \begin{align}
            E_i &= \cos^2\theta \ket{i}\bra{i}_S + \frac{\sin^2\theta}{d-1}\left(I-\ket{i}\bra{i}_S\right)\label{eq:povm_qudit}
        \end{align}   
        and the resulting measurement strength $\mathfrak{s}$ is:
        \begin{align}
            \mathfrak{s}=\frac{d\cos^2\theta-1}{d-1}\label{eq:strength_qudit}
        \end{align}
    \end{subequations}  
\end{example}

One can easily see that setting $\theta$ to $0$ and $\arccos \left(d^{-1/2}\right)$ yields the strong and weak case respectively. 
This particular choice of meter guarantees that the measurement be \emph{minimally disturbing} \cite{Wiseman2009a}, since the measurement operators (or Kraus operators) $M_i\equiv\prescript{}{M}{\bra{i}}U\ket{\phi}_M$ are directly equal in this case to the square root of the POVM effects,  $M_i=\sqrt{E_i}$. 
Also note that contrary to the traditionnal von Neumann measurement scheme \cite{vonNeumann1935}, the system-meter interaction used here is always \emph{strong}; the measurement strength is instead controlled by changing the initial meter state directly.
We shall adopt the same approach for our own construction.

\section{Two-outcome nonlocal VSM}
\label{sec:nvsm-1}

We now consider a multipartite quantum system $\ket{\psi}\in\left(\mathbb{C}^2\right)^{\otimes N}$ composed of $N$ qubits placed at different locations.

\begin{defn}
    We call a VSM \emph{nonlocal} if its strong limit PVM can be written as projectors on entangled states.
\end{defn}

The goal of this paper is to construct explicit nonlocal VSM implementations for multipartite qubit systems, focusing on PVMs associated to product observables.
In this Section, we first consider 2-outcome PVMs $\left\{P_+,P_-\right\}$, with $P_+$ and $P_-$ both of rank $2^{N-1}$.  

We first recall the constructions of \cite{Edamatsu2016a} and \cite{Vidil2019} to implement nonlocal measurements of bipartite spin products using a meter in an entangled state.
The key idea behind these is to rely on meter entanglement to couple only to the desired nonlocal properties of the system.

Here, we wish to extend this idea further, to the case of $N>2$ multipartite qubit systems.
In this situation, it is known that contrary to the bipartite case, there are several non-equivalent (in the sense of Stochastic Local Operations and Classical Communication) types of entanglement \cite{Dur2000,Bengtsson2017}.
One commonly used class of multipartite entangled states are the generalized Greenberger-Horne-Zeilinger (GHZ) states \cite{Greenberger1989,Bouwmeester1999}, the following two we shall use:
\begin{subequations}
    \begin{align}
        \ket{\text{GHZ}_N^+}&=\frac{1}{\sqrt{2}}\left(\ket{0}^{\otimes N}+\ket{1}^{\otimes N}\right)\label{eq:GHZ_def+}\\ 
        \ket{\text{GHZ}_N^-}&=\frac{1}{\sqrt{2}}\left(\ket{0}^{\otimes N}-\ket{1}^{\otimes N}\right)\label{eq:GHZ_def-} 
    \end{align}  
\end{subequations}

In \cite{Walck2008, Walck2009}, it has been shown that such states are the only ones capable of containing information at the $N$-partite level, as they remained undetermined by their reduced density matrices.
This makes them ideal candidates as nonlocal meters of a $N$-partite system. 
Indeed, one can directly see that tracing out any subsystem yields the same density matrix for both $\ket{\text{GHZ}_N^+}$ and $ \ket{\text{GHZ}_N^-}$, which makes them indistinguishable at the local level.
The GHZ states \eqref{eq:GHZ_def+} and \eqref{eq:GHZ_def-} are eigenstates of the product operator $X^{\otimes N}$, with eigenvalue $+1$ for $\ket{\text{GHZ}_N^+}$ and $-1$ for $\ket{\text{GHZ}_N^-}$.
This fact motivates us to use the meter observable $X^{\otimes N}$ as a register to encode nonlocal information about our system.

As an initial meter state, by analogy with \cite{Vidil2019}, we use:
\begin{align}
    \ket{\Phi(N,\theta)}=\cos\theta\ket{\text{GHZ}_N^+}+\sin\theta\ket{\text{GHZ}_N^-}
\end{align}
which we can express in the computational basis as:
\begin{align}
    \frac{1}{\sqrt{2}}\left\{\left(\cos\theta+\sin\theta\right)\ket{0}^{\otimes N}+\left(\cos\theta-\sin\theta\right)\ket{1}^{\otimes N}\right\}
\end{align}
We call such a state a \emph{nonlocal meter state} for $N$ qubits.

Suppose we wish to measure the PVM associated to the product observable $\mathcal{O}=\bigotimes_{n=1}^N\mathcal{O}_n$ where $\mathcal{O}_n$ is the Pauli operator $X$, $Y$ or $Z$ acting on the $n$-th subsystem alone.
As a coupling interaction between the system $\ket{\Psi}_S$ and the meter $\ket{\Phi}_M$, we apply a Controlled-$\mathcal{O}_n$ gate on the $n$-th system qubit, controlled by the $n$-th meter qubit.
The post-interaction total state is then:
\begin{equation}
    \begin{split}
        \ket{\Psi}_S\ket{\Phi(N,\theta)}_M\longrightarrow&\frac{1}{\sqrt{2}} \left\{ \left(\cos\theta+\sin\theta\right)\ket{\Psi}_S\ket{0}^{\otimes N}_M \right. \\
        &\left. {}+\left(\cos\theta-\sin\theta\right)\mathcal{O}\ket{\Psi}_S\ket{1}^{\otimes N}_M \right\} \label{eq:total_state_1}
    \end{split}    
\end{equation}
This use of qubit meters as control gates on the system being measured is similar to the method presented in \cite{Ogawa2019}.

We now aim to retrieve the information about the system by projectively measuring each part of the meter in the $X$-eigenbasis $\left\{\ket{+},\ket{-}\right\}$. 
To this end, we use the decompositions:
\begin{subequations}
    \label{eq:x_decomp}
    \begin{align}
        \ket{0}^{\otimes N} &= 2^{-N/2}\sum_{\underline{t}\in\left\{+,-\right\}^N}\ket{\underline{t}}\\
        \ket{1}^{\otimes N} &= 2^{-N/2}\sum_{\underline{t}\in\left\{+,-\right\}^N}\text{sgn}(\underline{t})\ket{\underline{t}}
    \end{align}
\end{subequations}
where $\underline{t}\equiv(t_1t_2\dots t_N)$ is a vector composed of $N$ symbols $+$ or $-$, and we define $\text{sgn}(\underline{t})=\prod_{k=1}^Nt_k$. 
For instance for $N=2$, we have $\text{sgn}(++)=\text{sgn}(--)=+1$ and $\text{sgn}(+-)=\text{sgn}(-+)=-1$.

Inserting the expressions \eqref{eq:x_decomp} in the state \eqref{eq:total_state_1}, we obtain:
\begin{equation}
    \begin{split}
        2^{-\frac{N+1}{2}}\sum_{\underline{t}\in\left\{+,-\right\}^N}&\Big{\{}\left(\cos\theta+\sin\theta\right)\ket{\Psi}_S \\
        &+\text{sgn}(\underline{t})\left(\cos\theta-\sin\theta\right)\mathcal{O}\ket{\Psi}_S\Big{\}}\ket{\underline{t}}_M \label{eq:final_state1}
    \end{split}
\end{equation}

From this, one can extract the measurement operator for each final meter state $M_{\underline{t}}$. 
Since the final state \eqref{eq:final_state1} only depends on the global sign of the meter $\text{sgn}(\underline{t})$, which is accessible to the observers once they classically communicate and multiply the outcomes of the local measurements, there are only 2 distincts measurement operators, each appearing $2^{N-1}$ times:
\begin{align}
    M_\pm= 2^{-\frac{N+1}{2}}\left\{\left(\cos\theta+\sin\theta\right)I\pm\left(\cos\theta-\sin\theta\right)\mathcal{O}\right\}
\end{align}

Using the spectral decomposition $\mathcal{O}=P_+-P_-$ and the completeness relation $I=P_++P_-$ we get:
\begin{align}
    M_\pm= 2^{-\frac{N-1}{2}}\left\{\cos\theta P_\pm+\sin\theta \left(I-P_\pm\right)\right\}
\end{align}

Knowing the measurement operators, one can directly compute the two POVM effects, taking the multiplicities into account:
\begin{align}
    E_\pm&= 2^{N-1} M_\pm^\dagger M_\pm=\cos^2\theta P_\pm+\sin^2\theta \left(I-P_\pm\right)\label{eq:povm2outcomes}
\end{align}

\begin{figure}[htp]
    \includegraphics[width=0.40\textwidth]{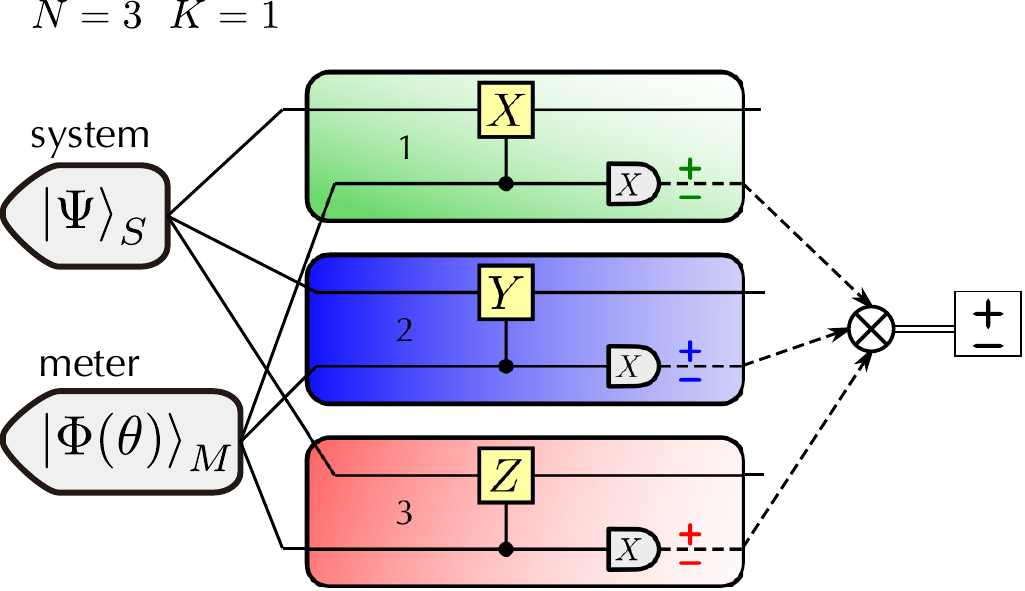}
    \caption{\label{fig:meas1}Example of an implementation of the two-outcome nonlocal VSM described in Sec. \ref{sec:nvsm-1} for the product $X_1Y_2Z_3$ when $N=3$.
     Entangled eigenstates remain unaffected in the strong measurement limit.}
\end{figure}

Comparing Eq. \eqref{eq:povm2outcomes} with Eq. \eqref{eq:povm_qudit}, it is apparent that the measurement protocol described here constitutes a minimally-disturbing implementation of a 2-outcome nonlocal VSM, with measurement strength $\mathfrak{s}=\cos2\theta$ similar to the one in \cite{Vidil2019}.
The nonlocal character is evident from the fact that $P_\pm$ admits entangled states in its image.

An example of this method applied to the observable $\mathcal{O}=X_1Y_2Z_3$ is depicted in \figref{fig:meas1}.
Entangled eigenstates such as $\frac{1}{\sqrt{2}}\left(\ket{+_z+_x+_y}+i\ket{-_z-_x-_y}\right)$ are unaffected by the measurement, as is expected from a genuine L{\"u}ders measurement \cite{Luders1951}.

\section{Many-outcome nonlocal VSM}
\label{sec:nvsm-2}

\begin{figure*}[htp]
    \includegraphics[width=0.80\textwidth]{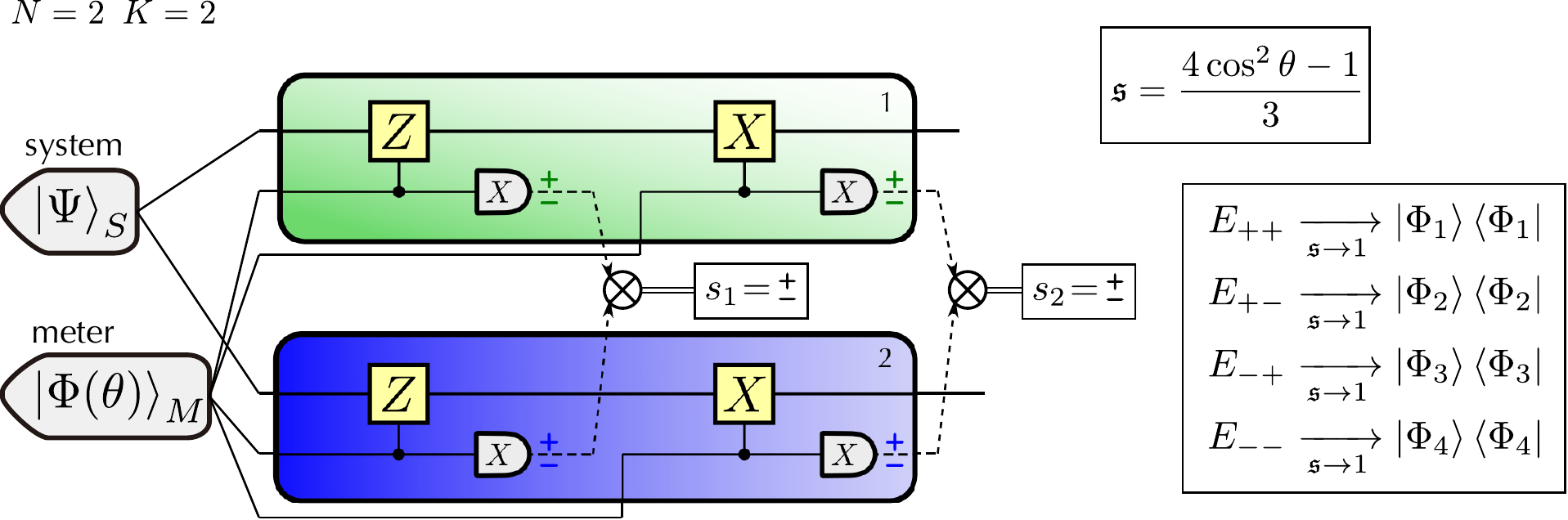}
    \caption{\label{fig:BellMeas}Schematic representation of a Variable Strength Bell State Measurement using the nonlocal meter $\ket{\Phi(2,2,\theta)}$ of \eqref{eq:meter_state_2}. 
    As the parameter $\theta$ approaches the value $0$, one is able to distinguish with high confidence between the 4 orthogonal Bell states looking at the measurement outcomes $s_1$ and $s_2$.
    When $\theta$ gets close to $\frac{\pi}{3}$, the measurement outcomes become random and uncorrelated to the input state.}
\end{figure*}

We now turn our attention to the case of a nonlocal measurement with more than two outcomes, again on a system of $N$ qubits. 
For $N=2$, an example of such a measurement that is of great importance for quantum information technologies, is the Bell state measurement \cite{Nielsen2010,Weinfurter1994} given by $\left\{\ket{\Phi_i}\bra{\Phi_i}\right\}_{i=1\dots4}$ where $\ket{\Phi_i}$ is the $i$-th Bell state. 
To our knowledge, a VSM implementation for this particular PVM has not yet been found.
In this Section, we shall propose such an implementation among other things.

We restrict ourselves to PVMs associated with joint measurements of $K$ commuting product observables $\mathcal{O}_1,\mathcal{O}_2,\dots,\mathcal{O}_K$, where each $\mathcal{O}_k$ is of the form $\bigotimes_{n=1}^N\mathcal{O}_{k,n}$ with $\mathcal{O}_{k,n}$ the Pauli operator $X$, $Y$ or $Z$ acting on the $n$-th subsystem alone.
For a system of $N$ qubits, there are at most $N$ independent such product observables that are pairwise commuting \cite{Ruan2004}, hence $K\leq N$.

Given that there are $2^K$ different outcomes, in analogy with Eq. \eqref{eq:qudit-meter} we use the following initial meter state, this time composed of $NK$ qubits: 
\begin{align}
    \begin{split}
        \ket{\Phi(K,N,\theta)}&=\frac{\sin\theta}{\sqrt{2^K-1}}\left\{\left(\ket{\text{GHZ}_N^+}+\ket{\text{GHZ}_N^-}\right)^{\otimes K}\right. \\
        &\left. {} -\ket{\text{GHZ}_N^+}^{\otimes K}\right\}+\cos\theta\ket{\text{GHZ}_N^+}^{\otimes K} \label{eq:meter_state_2}
    \end{split}    
\end{align}
Although complicated-looking at first glance, this expression simply shows that we assign an amplitude $\cos\theta$ to the one combination of GHZ states that yields the correct result, and give an equal weight to all other $2^K-1$ remaining combinations which lead to an incorrect result.
In the computational basis, this gives:
\begin{align}
    \begin{split}
        2^{-\frac{K}{2}}&\Big\{\left(\cos\theta+\sqrt{2^K-1}\sin\theta\right)\ket{0}^{\otimes NK}\\
        &+\left(\cos\theta-\frac{\sin\theta}{\sqrt{2^K-1}}\right)\sum_{\substack{\underline{l}\in\left\{0,1\right\}^K \\ \backslash\left\{0\right\}^K}}\bigotimes_{k=1}^K\ket{l_k}^{\otimes N}\Big\}\label{eq:meter_state_3}
    \end{split}  
    \raisetag{5\normalbaselineskip}  
\end{align}
where $\underline{l}\equiv(l_1\dots l_K)$ is a vector composed of $K$ symbols $0$ or $1$ that are not all simultaneously 0.
We call such a state a \emph{$K$-fold nonlocal meter state} for $N$ qubits.

To make the notation clearer, let us explicitly write down as an example the state $\ket{\Phi(2,3,\theta)}$:
\begin{align}
    \begin{split}
        &\frac{1}{2}\Big\{\left(\cos\theta+\sqrt{3}\sin\theta\right)\ket{000000}\\
        &+\left(\cos\theta-\frac{\sin\theta}{\sqrt{3}}\right)\left(\ket{111000}+\ket{000111}+\ket{111111}\right)\Big\}
    \end{split}  
    \raisetag{3.5\normalbaselineskip} 
\end{align}

Similarly to Sec. \ref{sec:nvsm-1}, we then couple this meter to a $N$-qubit system $\ket{\Psi}_S$ via a sequence of $K$ local Controlled-$\mathcal{O}_{k,n}$ ($k=1\dots K$) gates acting on each subsystem $n$.
Since we chose a set of $K$ commuting product operators, the global order of the coupling does not matter.
However, once an arbitrary order has been chosen, it must be respected for each subsystem.
The post-interaction total state is then, up to global factor $2^{-\frac{K}{2}}$:
\begin{align}
    \alpha\ket{\Psi}_S\ket{0}_M^{\otimes NK}+\beta\sum_{\substack{\underline{l}\in\left\{0,1\right\}^K \\ \backslash\left\{0\right\}^K}}\left(\prod_{k=1}^K\mathcal{O}_k^{l_k}\right)\ket{\Psi}_S\bigotimes_{k'=1}^K\ket{l_{k'}}_M^{\otimes N}\label{eq:total_state_2}
\end{align}
where we put for clarity $\alpha=\cos\theta+\sqrt{2^K-1}\sin\theta$ and $\beta=\cos\theta-\frac{\sin\theta}{\sqrt{2^K-1}}$.

We then rewrite this state in the $X$-basis as follows, by expressing each $\ket{l_k}_M^{\otimes N}$ as a combination of $\ket{\underline{t_k}}$ where $\underline{t_k}$ is a vector composed of $N$ symbols $+$ or $-$:
\begin{widetext}
    \begin{align}
       2^{-\frac{K(N+1)}{2}}\sum_{\underline{t_1}\in\left\{+,-\right\}^N}\dots\sum_{\underline{t_K}\in\left\{+,-\right\}^N}\Bigg(\alpha\ket{\Psi}_S+\beta\sum_{\substack{\underline{l}\in\left\{0,1\right\}^K \\ \backslash\left\{0\right\}^K}}\prod_{k=1}^K\left(\text{sgn}\left(\underline{t_k}\right)\mathcal{O}_k\right)^{l_k}\ket{\Psi}_S\Bigg)\ket{\underline{t_1}\dots\underline{t_K}}_M\label{eq:total_state_3}
    \end{align}
\end{widetext}

From Eq. \eqref{eq:total_state_3}, it is straightforward to extract $2^K$ different measurement operators, each appearing $2^{K(N-1)}$ times, by noticing that the final state only depends on the $K$ quantities $\text{sgn}\left(\underline{t_k}\right)$ which, for clarity's sake, we denote from now on by the vector $\underline{s}\equiv(s_1\dots s_K)$ with $s_k=\pm$.
The vector $\underline{s}$ of results is obtained in a similar way to that of Sec. \ref{sec:nvsm-1}, this time by classically combining the local measurement outcomes for each of the $K$ rounds separately:
\begin{align}
    M_{\underline{s}}=2^{-\frac{K(N+1)}{2}}\Big\{\alpha I +\beta\sum_{\substack{\underline{l}\in\left\{0,1\right\}^K \\ \backslash\left\{0\right\}^K}}\prod_{k=1}^K\left(s_k\mathcal{O}_k\right)^{l_k} \Big\} \label{eq:measop}
\end{align}
After inserting the joint PVM decompositions:
\begin{align}
    I&=\sum_{\underline{i}\in\left\{+,-\right\}^K}P_{i_1\dots i_K}\\
    \prod_{k=1}^K\left(s_k\mathcal{O}_k\right)^{l_k}&=\sum_{\underline{i}\in\left\{+,-\right\}^K}\prod_{k=1}^K\left(s_ki_k\right)^{l_k}P_{i_1\dots i_K}
\end{align} 
and some simplification, Eq. \eqref{eq:measop} becomes:
\begin{align}
    M_{\underline{s}}=2^{-\frac{K(N-1)}{2}}\left\{\cos\theta P_{\underline{s}} + \frac{\sin\theta}{\sqrt{2^K-1}}\left(I-P_{\underline{s}}\right)\right\}
\end{align}
From this, we obtain the POVM effects, taking into account the multiplicities:
\begin{align}
    E_{\underline{s}}=2^{K(N-1)}M_{\underline{s}}^\dagger M_{\underline{s}}=\cos^2\theta P_{\underline{s}} + \frac{\sin^2\theta}{2^K-1}\left(I-P_{\underline{s}}\right)
\end{align}

The POVM constructed here is therefore a nonlocal $2^K$-outcome VSM, with the measurement strength given by Eq. \eqref{eq:strength_qudit}, setting $d=2^K$:
\begin{align}
    \mathfrak{s}_K=\frac{2^K\cos^2\theta-1}{2^K-1}
\end{align}

Since when $K=N$, the projectors $P_{i_1\dots i_K}$ are all rank-1, the method presented here can be described as a Variable-Strength State Measurement.
An example of particular interest, due to its numerous applications in quantum information technologies, is the Bell State Measurement \cite{Nielsen2010,Weinfurter1994}.
Using the process represented on \figref{fig:BellMeas}, one can effectively discriminate between the four Bell states with a controllable measurement strength.

\section{Relation between measurement strength and meter tangle}
\label{sec:tangle}

Finally, we investigate the properties of the family of nonlocal meter states given by Eq. \eqref{eq:meter_state_2}. 
In particular, a relationship between the amount of meter entanglement and the measurement strength was established in \cite{Vidil2019} for the two-outcome two-qubit case ($N=2$, $K=1$ with our notation).
The measure of entanglement used was the concurrence \cite{Wootters1998}; however for simplicity we shall use here the square of the generalization of the concurrence, called the $n$-tangle $\tau_n$ \cite{Coffman2000, Wong2001}, defined for a $n$-partite quantum state by:
\begin{align}
    \begin{split}
        \tau_n=2&\Big|\sum a_{\alpha_1\dots\alpha_n}a_{\beta_1\dots\beta_n}a_{\gamma_1\dots\gamma_n}a_{\delta_1\dots\delta_n}\\
        &\times\epsilon_{\alpha_1\beta_1}\dots\epsilon_{\alpha_{n-1}\beta_{n-1}}\epsilon_{\gamma_1\delta_1}\dots\epsilon_{\gamma_{n-1}\delta_{n-1}}\\
        &\times\epsilon_{\alpha_n\gamma_n}\epsilon_{\beta_n\delta_n}\Big|
    \end{split}
    \label{eq:tangle}
\end{align}
where the components of the state in the computational basis are represented by a $n$th-order tensor $a$, $\epsilon_{ij}$ is the Levi-Civita symbol and the summation is over all the indices.
The $n$-tangle defined here has been shown to be an entanglement monotone only for $n=3$ and $n$ even, because it is not permutation-invariant for odd $n>3$ \cite{Wong2001}. 

The evaluation of $\tau_n$ via Eq. \eqref{eq:tangle} is usually computationally demanding for an arbitrary n-partite state.
However, the nonlocal meter state $\ket{\Phi(K,N,\theta)}$ given by Eq. \eqref{eq:meter_state_2} exhibits properties that allow us to calculate its $NK$-tangle directly.

Indeed, expanding the sum for the last indices and using the symmetries of the state, one can obtain the following simpler expression for $\tau_{NK}(\ket{\Phi})$:
\begin{align}
    \begin{split}
        \tau_{NK}(\ket{\Phi})=4&\Big(\sum a_{\alpha_1\dots\alpha_{(K-1)N}0\dots0}a_{\beta_1\dots\beta_{(K-1)N}1\dots1}\\
        &\times\epsilon_{\alpha_1\beta_1}\dots\epsilon_{\alpha_{(K-1)N}\beta_{(K-1)N}}\Big)^2
    \end{split}
\end{align}

A straightforward substitution and simplification leads to the following remarkable relation:
\begin{align}
    \tau_{NK}(\ket{\Phi(K,N,\theta)})=\left(\frac{2^K\cos^2\theta-1}{2^K-1}\right)^2=\mathfrak{s}_K^2
\end{align}
that is, the tangle of the nonlocal meter for $2^K$ outcomes is directly equal to the square of the resulting measurement strength.

\begin{figure}[htp]
    \includegraphics[width=0.40\textwidth]{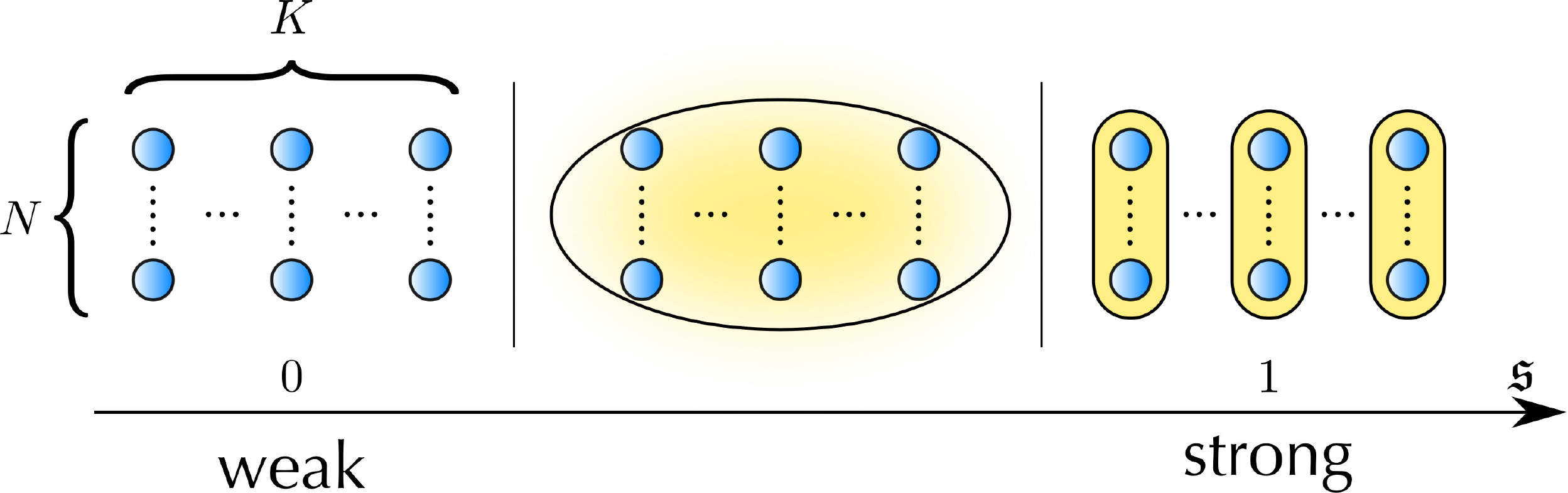}
    \caption{\label{fig:entanglement}Entanglement of the nonlocal qubit meter state $\ket{\Phi(K,N,\theta)}$ as a function of the measurement strength.
    The presence of entanglement, as measured by the $NK$-tangle $\tau_{NK}$, is represented here by a yellow coloring between qubits (in blue). 
    For $\mathfrak{s}=0$, the meter state is fully separable as a product of $NK$ qubits.
    For $\mathfrak{s}=1$, the meter state is separable as a product of $K$ GHZ states over $N$ qubits.
    Genuine $NK$-partite entanglement is only present at intermediate strengths $0<\mathfrak{s}<1$.
    }
\end{figure}

This generalizes the result of \cite{Vidil2019}, and allows us to interpret the $n$-tangle as a resource to perform nonlocal measurements: more $n$-tangled states are more reliable nonlocal meters.  
We also note that this relation still holds using an alternative measure of entanglement that is valid for odd numbers of parties, the odd $n$-tangle \cite{Li2012}.
Curiously, genuine NK-partite entanglement only appears for intermediate measurement strengths, as can be seen on \figref{fig:entanglement}.

While it is known that for $n\geq3$ there is no single measure of $n$-partite entanglement \cite{Bengtsson2017}, this relation offers a relatively simple operational interpretation for the $n$-tangle, which we hope will lead to further understanding of multipartite entanglement. 

Compared with other schemes that rely on quantum teleportation or entanglement swapping \cite{Vaidman2003,Brodutch2016}, and thus always require a meter in a maximally-entangled state regardless of the measurement strength, the scheme presented here is resource-efficient in terms of meter entanglement.
This offers an advantage in terms of implementation, as multipartite entanglement is notoriously hard to generate and maintain over a significant period of time \cite{Pu2018,Li2020}.
This also allows for a quantitative interpretation of results obtained with imperfect entanglement sources used as meters.

\section{Conclusions}
\label{sec:conclusion}
In this paper, we defined a variable-strength measurement as a measurement whose POVM linearly depends on a single real parameter, which allows the experimenter to continuously change the measurement behavior from strong measurement to no measurement.
We explicitly constructed an indirect measurement model to realize such measurements in the case of nonlocal multipartite systems, for which spatial separation prevent any kind of direct von Neumann-like coupling.
To do so, we introduced a set of nonlocal meter states, which are to be coupled locally to the system, so that the classically combined measurement outcomes yield the correct global result.
In the case of product operators with 2 outcomes, we proved that our model reproduces in a minimally-disturbing manner the expected statistics of a nonlocal VSM.
Moreover, we showed that by successively coupling to product operators that commute, our scheme can be extended to the $2^K$-outcome case, acting as a variable-resolution nonlocal state measurement in the case $K=N$.
In this construction, the measurement strength is directly given by the square root of the tangle of the nonlocal meter state, a generalization of the result of \cite{Vidil2019} which establishes the $n$-tangle as a resource for nonlocal generalized measurements.

This scheme should be implementable using linear optics with entangled photon pairs in the 2-outcome case. 
For the general many-outcome case, e.g. the variable-strength Bell state measurement, the challenge resides in the generation of the appropriate 4-qubit entangled meter state.

Nonlocal VSM can be used for the measurement of nonlocal weak values, which have both fundamental and practical applications \cite{Brodutch2016}.
We expect this work to open as well the door to experimental studies of error-disturbance relations \cite{Edamatsu2016b} in the case of sequential nonlocal measurements, beyond the simpler bipartite case.
\begin{acknowledgments}
    The authors thank F. Kaneda and S. Baek for helpful remarks and discussions. 
    This research was supported by the MEXT Quantum Leap Flagship Program (MEXT Q-LEAP) Grant Number JPMXS0118067581. 
\end{acknowledgments}


%

\end{document}